\documentclass[10pt, oneside,twocolumn,a4paper]{article}
\usepackage{graphicx}
\usepackage[utf8]{inputenc}
\usepackage{mathptmx}
\usepackage{tabularx}
\usepackage{ragged2e}
\usepackage[singlelinecheck=false]{caption}
\usepackage[T1]{fontenc}
\usepackage[numbers,sort&compress]{natbib}
\usepackage{hyperref}
\usepackage{amsmath}
\RequirePackage[blocks]{authblk}
\usepackage{babel}
\usepackage{color}
\usepackage{amssymb}

\usepackage{graphicx}
\usepackage{float}
\usepackage{blindtext}
\usepackage{subcaption}

\makeatletter
\let\ps@plain\ps@empty
\def\@xivpt{14bp}

\def\@sect#1#2#3#4#5#6[#7]#8{%
  \ifnum #2>\c@secnumdepth
    \let\@svsec\@empty
  \else
    \refstepcounter{#1}%
    \protected@edef\@svsec{%
      \ifnum #2<4
        \hb@xt@10mm{\csname the#1\endcsname}\relax
      \else
        \hb@xt@12mm{\csname the#1\endcsname}\relax
      \fi}%
  \fi
  \@tempskipa #5\relax
  \ifdim \@tempskipa>\z@
    \begingroup
      #6{%
        \@hangfrom{\hskip #3\relax\@svsec}%
          \interlinepenalty \@M #8\@@par}%
    \endgroup
    \csname #1mark\endcsname{#7}%
    \addcontentsline{toc}{#1}{%
      \ifnum #2>\c@secnumdepth \else
        \protect\numberline{\csname the#1\endcsname}%
      \fi
      #7}%
  \else
    \def\@svsechd{%
      #6{\hskip #3\relax
      \@svsec #8}%
      \csname #1mark\endcsname{#7}%
      \addcontentsline{toc}{#1}{%
        \ifnum #2>\c@secnumdepth \else
          \protect\numberline{\csname the#1\endcsname}%
        \fi
        #7}}%
  \fi
  \@xsect{#5}}
\renewcommand\LARGE{\@setfontsize\LARGE{16}{20}}
\def\abstract#1{\def\@abstract{#1}}
\def\abstractEn#1{\def\@abstractEn{#1}}
\def\titleEn#1{\def\@titleEn{#1}}
\def\@maketitle{%
  \newpage
  \null
  \let \footnote \thanks
    {\LARGE\bfseries\RaggedRight \@title \par}%
    \vskip 1\baselineskip%
    {\normalsize
      \@author\par}%
    \vskip 2\baselineskip%
    {\section*{Abstract}
     \@abstract}%
        \par
        \vskip 1\baselineskip
      \keywords{Redispatch, Grid Control Optimization, Security Constrained Optimal Power Flow, Sequential Quadratic Constrained Quadratic Programming, Particle Swarm Optimization}
  \par
  \vskip 2\baselineskip}

\renewcommand\section{\@startsection {section}{1}{\z@}%
                                   {-3.5ex \@plus -1ex \@minus -.2ex}%
                                   {\baselineskip}%
                                   {\normalfont\Large\bfseries\RaggedRight}}
\renewcommand\subsection{\@startsection{subsection}{2}{\z@}%
                                     {\baselineskip}%
                                     {1ex}%
                                     {\normalfont\large\bfseries\RaggedRight}}
\renewcommand\subsubsection{\@startsection{subsubsection}{3}{\z@}%
                                     {1\baselineskip}%
                                     {3bp}%
                                     {\normalfont\normalsize\bfseries\RaggedRight}}
\renewcommand\paragraph{\@startsection{paragraph}{4}{\z@}%
                                    {1\baselineskip\@plus1ex \@minus.2ex}%
                                    {3bp}%
                                    {\normalfont\normalsize\RaggedRight}}
\renewcommand\subparagraph{\@startsection{subparagraph}{5}{\parindent}%
                                       {3.25ex \@plus1ex \@minus .2ex}%
                                       {-1em}%
                                      {\normalfont\normalsize\bfseries\RaggedRight}}
\parindent\p@
\makeatother
\DeclareCaptionLabelSeparator{enskip}{\enskip}
\providecommand{\keywords}[1]{\textbf{\textit{Keywords---}} #1}

\title{Comparison of Convexificated SQCQP and PSO for the \\ Optimal Transmission System Operation based on Incremental \\ In-Phase and Quadrature Voltage Controlled Transformers}
\author[1]{Marcel Sarstedt}	
\author[1]{Thomas Leveringhaus}
\author[1]{Leonard Kluß}
\author[1]{Lutz Hofmann}
\affil[1]{Institute of Electric Power Systems, Electric Power Engineering Section, Leibniz Universität Hannover, Germany}
\affil[1]{[surname]@ifes.uni-hannover.de}

\abstract{The optimal operation of electrical energy systems by solving a security constrained optimal power flow (SCOPF) problem is still a challenging research aspect. Especially, for conventional optimization methods like sequential quadratic constrained quadratic programming (SQCQP) the formulation of the incremental control variables like in-phase and quadrature voltage controlled transformers in a solver suitable way is complex. Compared to this, the implementation of these control variables within heuristic approaches like the particle swarm optimization (PSO) is simple but problem specific adaptations of the classic PSO algorithm are necessary to avoid an unfortunate swarm behavior and local convergence in bad results. The objective of this paper is to introduce a SQCQP and a modified PSO approach in detail to solve the SCOPF problem adequately under consideration of flexible incremental in-phase and quadrature transformers tap sets and to compare and benchmark the results of both approaches for an adapted IEEE 118-bus system. The case-study shows that both approaches lead to suitable results of the SCOPF with individual advantages of the SQCQP concerning the quality and the reproducibility of the results while the PSO lead to faster solutions when the complexity of the investigation scenario increases.}

\begin{document}
\hypersetup{
pdfproducer={}}
\maketitle

\section{Motivation}
The massive integration of decentral energy resources especially to the distribution grid level leads to a transition of the electric power system. The share of conventional thermal power plants in the energy mix is reduced because of the power supply priority of renewables and the phase-out of nuclear and fossil fuel based generation in Germany. The former unidirectional active and reactive power flows from the transmission to the distribution grid level become bidirectional and volatile and the directions of active and reactive power flows are decoupled. The existing grid is more stressed in operating points with a high share of renewables and partially critical system states may arise due to long distances between power generation and loads. A variety of grid control actions by the system operators are necessary to guarantee a secure and reliable energy supply. The distribution grid level becomes increasingly active due to the flexibility potentials of converter coupled energy resources, the massive integration of measurement infrastructure and information and communication technologies. In contrast to that, the flexibility potentials regarding the provision of ancillary services for the grid control decrease at the transmission grid level. Cost-intensive grid expansion measures and a large number of redispatch actions by the transmission system operators are the consequence. The optimal planning of these redispatch actions and additional grid control measures~(e.g.~transformer~tap~sets) focusing a grid loss reduction in the interest of common welfare can be described by an security constrained optimal power flow problem (SCOPF) \cite{carpentier1962contribution,4073461}. For the solution of the SCOPF suitable robust optimization methods are necessary that can deal with the specific requirements of the meshed transmission grid level and a variety of different flexibilities to avoid local convergence. Operational degrees of freedom within this paper are represented by the active and reactive power redispatch and voltage control of thermal power plants, the curtailment of renewable power supply as well as incremental in-phase and quadrature voltage control of transformers.

Solving a SCOPF is challenging, due to its non\hbox{-}convexity, np-hardness  \cite{bienstock2019strong} and high number of degrees of freedom~\cite{frank2012optimal}. Therefore, this paper presents the comparison of two, independently developed optimization methods (see section \ref{AdaptationGeneral}) for the solution of the SCOPF problem described in section~\ref{DefOPF}. The first approach is a solution by a sequentially solved convexificated Quadratically Constrained Quadratic Program (SQCQP) based on \cite{7286314,leveringhaus_optimal_2016,8601613,Lev1}, which is presented in section \ref{AdaptationSQCQP}. The second approach solves a mixed integer, non-linear optimization problem using a modified particle swarm optimization (PSO) e.g. with an enhanced velocity control of the swarm (see section \ref{AdaptationPSO}, \cite{Sar1}). For the evaluation of the swarm fitness in the PSO as well as of the quality of approximation in the SQCQP a Newton Raphson power flow calculation is used. The focus of the PSO approach is a robust application to the SCOPF problem by avoiding local optima without an optimization of the PSO hyperparameters. The intention of the comparison and the main contribution of this paper are the benchmark and reproducibility of both approaches for the solution of SCOPF based problems at the transmission grid level and the identification of individual advantages (e.g. computation time, quality of the results). The benchmark scenario is based on an adaption of the IEEE 118-bus transmission grid regarding German transmission grid characteristic, introduced in \cite{Aachen}. A MathWorks MATLAB dataset of the grid and the results of the case study will be accessible at \cite{DATASET} for reproducibility reasons. The scenarios for the case-study and the results for the comparison of both approaches are presented in section~\ref{CaseStudiesGeneral}.    

\section{Definition of the SCOPF Problem}\label{DefOPF}
In general the formulation of the SCOPF problem within this paper is based on \cite{frank2012optimal,4073461}. The first summand of the SCOPF objective function (see Eq.~\ref{eq:SCOPF_obj}) to be minimized considers the grid losses $P_{\mathrm{loss}}$ monetized with the cost factor $c_{\mathrm{loss}}$. The second summand is the monetarization of nodal active power adaptations~$\mathrm{\Delta}\pmb{p}_{\mathrm{N}}$ using the energy resource specific redispatch cost vector~$\pmb{c}_{\mathrm{N}}$. The grid losses are a function of the complex nodal voltages $\underline{\pmb{v}}_{\mathrm{N}}$: 
\begin{align}\label{eq:SCOPF_obj}
\mathrm{min}\left(c_{\mathrm{loss}} \cdot P_{\mathrm{loss}}\left(\underline{\pmb{v}}_{\mathrm{N}}\right)+\pmb{c}_{\mathrm{N}}^{\mathrm{T}} \cdot \mathrm{\Delta}\pmb{p}_{\mathrm{N}}\right)
\end{align}
s.t.
\begin{align}
\pmb{v}_{\mathrm{N,min}} \leq \pmb{v}_{\mathrm{N}} \left(\underline{\pmb{v}}_{\mathrm{N}}\right) \leq \pmb{v}_{\mathrm{N,max}} \label{eq:Constraints1} \\
\pmb{i}_{\mathrm{T}}\left(\underline{\pmb{v}}_{\mathrm{N}}\right) \leq \pmb{i}_{\mathrm{T,max}} \label{eq:Constraints2}\\
\left(\pmb{p}_{\mathrm{N0}} + \mathrm{\Delta}\pmb{p}_{\mathrm{N}}\right) + \mathrm{j}\left(\pmb{q}_{\mathrm{N0}} + \mathrm{\Delta}\pmb{q}_{\mathrm{N}}\right) = 3\cdot\underline{\pmb{V}}_{\mathrm{N}}\cdot\underline{\pmb{Y}}_{\mathrm{NN}}^{*}\cdot\underline{\pmb{v}}_{\mathrm{N}}^{*} \label{eq:Constraints3}
\end{align}
The constraints (see Eq. \ref{eq:Constraints1}, \ref{eq:Constraints2} and \ref{eq:Constraints3}) of the SCOPF are represented by the minimum and maximum of the absolute value of the nodal voltages ($\pmb{v}_{\mathrm{N,min}}$, $\pmb{v}_{\mathrm{N,max}}$), the maximum admissible absolute value of the terminal current of the lines and transformers ($\pmb{i}_{\mathrm{T,max}}$) and the balances of nodal active and reactive power ($\pmb{p}_{\mathrm{N}}$,~$\pmb{q}_{\mathrm{N}}$) within the power equation of Eq.~\ref{eq:Constraints3}. Thereby, $\underline{\pmb{V}}_{\mathrm{N}}$ is the diagonal matrix of the nodal voltages~$\underline{\pmb{v}}_{\mathrm{N}}$ and $\underline{\pmb{Y}}_{\mathrm{NN}}$ is the nodal admittance matrix of the system. 

\begin{figure}[h]
    \centering
    \includegraphics[width=235pt]{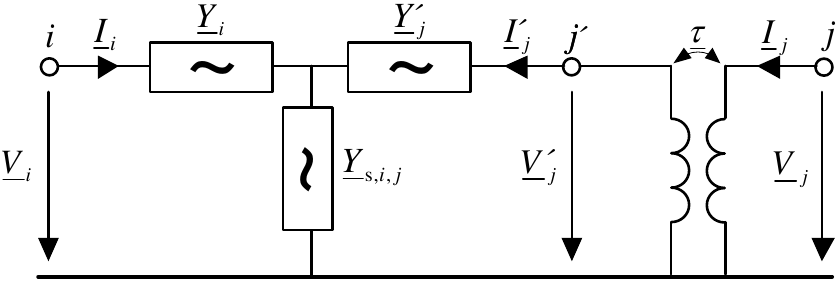}
    \caption{T-equivalent circuit of a transformer}
    \label{fig:TCircuit}
\end{figure}

The first two categories of operational degrees of freedom are redispatch measures represented by the change of the active and reactive power supply ($\mathrm{\Delta}\pmb{p}_{\mathrm{N}}$, $\mathrm{\Delta}\pmb{q}_{\mathrm{N}}$) of the different energy resources within the limits: 
\begin{align}\label{eq:Constraints4}
\mathrm{\Delta}\pmb{p}_{\mathrm{N,min}} \leq \mathrm{\Delta}\pmb{p}_{\mathrm{N}}  \leq \mathrm{\Delta}\pmb{p}_{\mathrm{N,max}} \\
\mathrm{\Delta}\pmb{q}_{\mathrm{N,min}} \leq \mathrm{\Delta}\pmb{q}_{\mathrm{N}}  \leq \mathrm{\Delta}\pmb{q}_{\mathrm{N,max}} \label{eq:Constraints5} 
\end{align}
At the slack and generation buses the voltage magnitudes are control variables. The flexibility range for the voltage control are specified by Eq. \ref{eq:Constraints1} and  constraints of the SCOPF are given by Eq. \ref{eq:Constraints5}. Further  operational degrees of freedom originate from the flexible adaptation of tap changer positions regarding an in-phase and quadrature voltage control of the transformers at their lower voltage side $j$. The implementation of both is based on the general \mbox{T-equivalent} circuit of a transformer (see Fig. \ref{fig:TCircuit}).  
The electric behavior at the higher voltage level node $i$ and the lower voltage level node $j$ is described by the terminal currents ($\underline{I}_{i},\underline{I}_{j}$), the terminal voltages ($\underline{V}_{i},\underline{V}_{j}$) interconnected by the terminal admittance matrix $\underline{\pmb{Y}}_{ij}$ within the two-port equation:
\begin{align}\label{eq:transformer1}
	\begin{bmatrix}
\underline{I}_{i} \\
\underline{I}_{j}  
\end{bmatrix} & = \underline{\pmb{Y}}_{ij}
	\begin{bmatrix}
	\underline{V}_{i} \\
\underline{V}_{j}  
\end{bmatrix} \\
 \underline{\pmb{Y}}_{ij} & =  \frac{1}{\underline{Y}_{i}+\underline{Y}_{j}^\prime+\underline{Y}_{\mathrm{s},i,j}} \begin{scriptsize} 
\begin{bmatrix}
\underline{Y}_{i}\left(\underline{Y}_{j}^\prime+\underline{Y}_{\mathrm{s},i,j} \right) & -\underline{\tau}_{j} \ \underline{Y}_{i} \ \underline{Y}_{j}^\prime\\
-\underline{\tau}_{j}^{*} \ \underline{Y}_{i} \ \underline{Y}_{j}^\prime & \left|\underline{\tau}_{j}\right|^{2} \ \underline{Y}_{j}^\prime\left(\underline{Y}_{i}+\underline{Y}_{\mathrm{s},i,j} \right)
\end{bmatrix} \end{scriptsize}
\label{eq:transformer2}
\end{align}
The rated terminal admittance matrix $\underline{\pmb{Y}}_{ij,\mathrm{r}}$ without in-phase and quadrature voltage control is based on Eq. \ref{eq:ratio}. The rated phase shifting is specified by the vector group code number~$k$ (e.g. $k=5$ for vector group Yd5):
\begin{align}\label{eq:ratio}
\underline{\tau}_{j,\mathrm{r}} = \frac{V_{i,\mathrm{r}}}{V_{j,\mathrm{r}}}\mathrm{e}^{\mathrm{j}k30^{\circ}} \ \mathrm{with} \ V_{i,\mathrm{r}}>V_{j,\mathrm{r}}
\end{align}
The adaptation of $\underline{\pmb{Y}}_{ij,\mathrm{r}}$ in case of in-phase and quadrature voltage control is described with the help of a diagonal tap changing matrix $\underline{\pmb{T}}_{ij,\mathrm{rel}}$ and the relative tap changings $\underline{\tau}_{i,\mathrm{rel}}$ and $\underline{\tau}_{j,\mathrm{rel}}$ in Eq. \ref{eq:TCMatrix}, wheras $\underline{\tau}_{i,\mathrm{rel}}$ typically equals 1 because only one side of the transformer has tap changers. 
\begin{align}\label{eq:TCMatrix}
\underline{\pmb{T}}_{ij,\mathrm{rel}}=\begin{bmatrix}
     \underline{\tau}_{i,\mathrm{rel}}=1      &  0\\
    0       & \underline{\tau}_{j,\mathrm{rel}}
\end{bmatrix}
\end{align}
Thereby, the in-phase voltage control is specified by the voltage change integer $n$ and the increment of the relative voltage change $\Delta V_{j,\mathrm{inc}}$ at the lower voltage side. The quadrature voltage control is given by the phase shift integer $m$ and the increment of the phase shift $\Delta \phi _{j,\mathrm{inc}}$. Both are combined in Eq. \ref{eq:tau} and reformulated in Eq. \ref{eq:tau2}, based on the transformer model in \cite{OptiChallenge}:
\begin{align}\label{eq:tau}
\underline{\tau}_{j,\mathrm{rel}} = \frac{\mathrm{e}^{\mathrm{j} m_{j} \Delta \phi_{j,\mathrm{inc}} }}{1+n_{j}\Delta V_{j,\mathrm{inc}}} \\
\underline{\tau}_{j,\mathrm{rel}}\cdot({1+n_{j}\Delta V_{j,\mathrm{inc}}}) = \mathrm{e}^{\mathrm{j} m_{j} \Delta \phi_{j,\mathrm{inc}}} \label{eq:tau2}
\end{align}
The increment of the relative voltage change $\Delta V_{j,\mathrm{inc}}$ and the increment of the phase shift $\Delta \phi _{j,\mathrm{inc}}$ are constant parameters of a specific transformer type. The additional non-convexity resulting from the integer property of the transformer steps complicates the solution process. The limits of the in-phase and quadrature voltage control of each transformer are given by the  minimum and maximum phase shift and voltage change integers, respectively:
\begin{align}\label{eq:ConstraintsTap1}
m_{j,\mathrm{min}} \leq m_{j}  \leq m_{\mathrm{j,max}}, \ m_{j} \in \mathbb{Z} \\
n_{j,\mathrm{j,min}} \leq n_{j}  \leq n_{\mathrm{j,max}} \ n_{j} \in \mathbb{Z} \label{eq:ConstraintsTap2}
\end{align}
The resulting terminal admittance matrix $\underline{\pmb{Y}}_{ij}$ of a transformer can be calculated with Eq. \ref{eq:YijTC}. 
\begin{align}\label{eq:YijTC}
\underline{\pmb{Y}}_{ij} = \underline{\pmb{T}}_{ij,\mathrm{rel}}^{*}\cdot\underline{\pmb{Y}}_{ij,\mathrm{r}}\cdot\underline{\pmb{T}}_{ij,\mathrm{rel}}
\end{align}
To combine the two-port equations of all transmission assets, the two-port matrices $\underline{\pmb{Y}}_{ij,\mathrm{r}}$ are arranged in a block diagonal matrix $\underline{\pmb{Y}}_{\mathrm{TT,r}}$ and the tap changing matrices $\underline{\pmb{T}}_{ij,\mathrm{rel}}$ are arranged in a block diagonal matrix $\underline{\pmb{T}}_{\mathrm{T,rel}}$. So in Eq.~\ref{eq:YTT} the matrix form of Eq. \ref{eq:YijTC} and the terminal admittance matrix $\underline{\pmb{Y}}_{\mathrm{TT}}$ result.
In Eq. \ref{eq:TAU2} the additionally necessary matrix form of Eq. \ref{eq:tau2} is given, in which $\textbf{e}$ is a one vector, $\pmb{M}_{\mathrm{T}}$ and $\pmb{N}_{\mathrm{T}}$ are the diagonal matrices of all $m_{j}$ and $n_{j}$ and $\pmb{\Delta v}_{\mathrm{T,inc}}$ and $\pmb{\Delta \varphi}_{\mathrm{T,inc}}$ are vectors of the increment parameters $\Delta V_{j,\mathrm{inc}}$ and  $\Delta \phi _{j,\mathrm{inc}}$:
\begin{align}\label{eq:YTT}
\underline{\pmb{Y}}_{\mathrm{TT}} = \underline{\pmb{T}}_{\mathrm{T,rel}}^{*}\cdot\underline{\pmb{Y}}_{\mathrm{TT,r}}\cdot\underline{\pmb{T}}_{\mathrm{T,rel}} \\
\underline{\pmb{T}}_{\mathrm{T,rel}}\cdot({\pmb{1}_{\mathrm{T}}+\pmb{N}_{\mathrm{T}}\cdot\pmb{\Delta v}_{\mathrm{T,inc}}}) = \mathrm{e}^{\mathrm{j} \pmb{M}_{\mathrm{T}}\cdot\pmb{\Delta \varphi}_{\mathrm{T,inc}}} \label{eq:TAU2}
\end{align}
For transmission lines and transformers terminals without voltage control, the matrix $\underline{\pmb{T}}_{\mathrm{T,rel}}$ only contains ones on its corresponding diagonal elements and: 
\begin{align}
m_{j,\mathrm{min}}=m_{j,\mathrm{max}}=n_{j,\mathrm{min}}=n_{j,\mathrm{max}}=0\\
\Delta V_{j,\mathrm{inc}}=0\\
\Delta \phi _{j,\mathrm{inc}}=0
\end{align}
To consider the topology of the network the node-terminal-incidence matrix $\pmb{I}_{\mathrm{NT}}$ is used to calculate the nodal admittance matrix $\underline{\pmb{Y}}_{\mathrm{NN}}$ in Eq. \ref{eq:YNN}. The incidence matrix $\pmb{I}_{\mathrm{NT}}$ describes the topology of the grid by the specification of the connection nodes $i$ and $j$ of each two-port:
\begin{align}\label{eq:YNN}
\underline{\pmb{Y}}_{\mathrm{NN}}= -\pmb{I}_{\mathrm{NT}}
\cdot\underline{\pmb{Y}}_{\mathrm{TT}}
\cdot\pmb{I}_{\mathrm{NT}}^{\mathrm{T}} 
\end{align}
Eq. \ref{eq:YTT}, \ref{eq:TAU2} and \ref{eq:YNN} need to be added to the SCOPF in Eq. \ref{eq:SCOPF_obj}-\ref{eq:Constraints3} to consider the additional degrees of freedom of transformer tap changings.
\section{Optimization Methods for the Solution of the SCOPF}\label{AdaptationGeneral}
The resulting formulation of the SCOPF has been derived in the previous section. As described in section 1, this paper presents the comparison of  two,  independently  developed optimization methods for the solution of the SCOPF problem. Adaptions of the problem formulation according to specific needs of the approach and adaptions of solver parameters are described in the following sections.
\subsection{SQCQP Approach}\label{AdaptationSQCQP}
The first approach applies a sequentially solved novel convexificated Quadratically Constrained Quadratic Program (SQCQP) based on \cite{7286314,leveringhaus_optimal_2016,8601613,Lev1} that is further developed in this paper to consider in-phase and quadrature voltage controlled transformers. For the convexification, the equations of the SCOPF need to be reformulated as real-valued system of equations with equal number of equations and variables and a maximal polynomial degree of two. Therefor, auxiliary equations and variables are implemented. In the following these auxiliary variables will have an additional index $a$. All equations are split into their real and imaginary part in Cartesian coordinates. 
Eq. \ref{eq:TAU2} takes a special role, because if proceeding in the aforementioned way, the variables $m_{j}$ would be part of the argument of sine and cosine functions, and therefore there would be non-quadratic functions.
Due to the integer property of $m_{j}$, exact reformulations as piecewise linear functions with vertices at all integer values of $m_{j}$ are a theoretical option, but the Special-Ordered-Sets required for these approximations would massively slow down the solution process of each sequential step and massively extend the memory requirements.
Approximating these functions with first or second-order Taylor series approximations in each sequential step of the SQCQP has been tested but led to high approximation deviations and bad convergence.
Deeper research has led to a somewhat counterintuitive way, but that has proven to be advantageous: Eq. \ref{eq:TAU2} is not to be split into its real and imaginary parts in Cartesian coordinates, but into its squared absolute values and into the tangents of its angles, whereas the transformer tap changes in $\underline{\pmb{T}}_{\mathrm{T,rel}}$ are still to be split into their real parts $\pmb{T}_{\mathrm{T,rel,r}}$ and imaginary parts~$\pmb{T}_{\mathrm{T,rel,i}}$ (likewise for the vector $\underline{\pmb{\tau}}_{\mathrm{T,rel}}$). This procedure does not yet require approximations at this stage and so Eq. \ref{eq:absT} results for the squared absolute values and Eq. \ref{eq:tanT} results for the tangents (whereas $\pmb{t}_{\mathrm{T}}\left(\pmb{m}_{\mathrm{T}}\right)$ is an abbreviation for the tangents):
\begin{align}\label{eq:absT}
\left(\pmb{T}_{\mathrm{T,rel,r}}^2+\pmb{T}_{\mathrm{T,rel,i}}^2\right)\cdot({\pmb{1}_{\mathrm{T}}+\pmb{N}_{\mathrm{T}}\cdot\pmb{\Delta v}_{\mathrm{T,inc}}})^2 = \pmb{1}_{\mathrm{T}}^2 \\
\frac{\pmb{\tau}_{\mathrm{T,rel,i}}}{\pmb{\tau}_{\mathrm{T,rel,r}}} = \tan\left({\pmb{M}_{\mathrm{T}}\cdot\pmb{\Delta \varphi}_{\mathrm{T,inc}}}\right)=\pmb{t}_{\mathrm{T}}\left(\pmb{m}_{\mathrm{T}}\right)
\label{eq:tanT}\end{align}
Eq \ref{eq:absT} is of fourth degree but can be reformulated into two equations of second degree polynomial. The result is shown in Eq. \ref{eq:absT1} and \ref{eq:absT2}.
\begin{align}\label{eq:absT1}
\pmb{\tau}_{\mathrm{T,rel,r}}^2+\pmb{\tau}_{\mathrm{T,rel,i}}^2 = \pmb{\tau}_{\mathrm{T,rel}}^2 \\
\pmb{T}_{\mathrm{T,rel}}\cdot({\pmb{1}_{\mathrm{T}}+\pmb{N}_{\mathrm{T}}\cdot\pmb{\Delta v}_{\mathrm{T,inc}}}) = \pmb{1}_{\mathrm{T}} \label{eq:absT2}
\end{align}
The geometry of the function in Eq \ref{eq:tanT} has been investigated more in detail and the tangent function can be well approximated for small angles - and only those are relevant at quadrature voltage control - by a linearization. It should be emphasized that thus approximations are used in each step of the SQCQP. To enable convergence, the point of linearization must be updated in each sequential step $\nu$. The result is shown in Eq. \ref{eq:tanTap}. The overall system of equations is shown in  Eq. \ref{eq:SOE} and \ref{eq:SOE2}.
\begin{align}\label{eq:tanTap}
\pmb{\tau}_{\mathrm{T,rel,i}}-\pmb{\tau}_{\mathrm{T,rel,r}}\left(\pmb{t}_{\mathrm{T}}\left(\pmb{m}_{\mathrm{T}}\right)\vert_{\nu}+\frac{\partial\pmb{t}\left(\pmb{m}_{\mathrm{T}}\right)}{\partial\pmb{m}_{\mathrm{T}}^{\mathrm{T}}}\vert_{\nu}\cdot\Delta\pmb{m}_{\mathrm{T}}\right)\approx\pmb{0}_{\mathrm{T}}
\end{align}
\begin{align}\label{eq:SOE}
\begin{small}
\begin{bmatrix}
    \pmb{0}_{\mathrm{T}}\\
    \pmb{1}_{\mathrm{T}}\\
    \pmb{0}_{\mathrm{T}}\\
    \pmb{m}_{\mathrm{T}}\\
    \pmb{n}_{\mathrm{T}}\\
    \pmb{0}_{\mathrm{T}}\\
    \pmb{0}_{\mathrm{T}}\\
    \pmb{0}_{\mathrm{T}}\\
    \pmb{0}_{\mathrm{T}}\\
    \pmb{0}_{\mathrm{T}}\\
    \pmb{0}_{\mathrm{T}}\\
    \pmb{0}_{\mathrm{T}}\\
    \mathrm{\Delta}\pmb{p}_{\mathrm{N}}\\
    \mathrm{\Delta}\pmb{q}_{\mathrm{N}}\\
    \pmb{0}_{\mathrm{N}}\\
    0
\end{bmatrix}
=
\pmb{f}\left(\begin{bmatrix}
    \pmb{\tau}_{\mathrm{T,rel,r}}\\
    \pmb{\tau}_{\mathrm{T,rel,i}}\\
    \pmb{\tau}_{\mathrm{T,rel}}\\
    \pmb{m}_{\mathrm{T}}\\
    \pmb{n}_{\mathrm{T}}\\
    \pmb{u}_{\mathrm{T,a,r}}\\
    \pmb{u}_{\mathrm{T,a,i}}\\
    \pmb{i}_{\mathrm{T,a,r}}\\
    \pmb{i}_{\mathrm{T,a,i}}\\
    \pmb{i}_{\mathrm{T,r}}\\
    \pmb{i}_{\mathrm{T,i}}\\
    \pmb{i}_{\mathrm{T}}\\
    \pmb{u}_{\mathrm{N,r}}\\
    \pmb{u}_{\mathrm{N,i}}\\
    \pmb{u}_{\mathrm{N}}\\
    P_{\mathrm{loss}}
\end{bmatrix}\right) 
\end{small}
\end{align}
\begin{align}
\begin{small}
=
\label{eq:SOE2}
\begin{bmatrix}
    \pmb{\tau}_{\mathrm{T,rel,r}}^2+\pmb{\tau}_{\mathrm{T,rel,i}}^2-\pmb{\tau}_{\mathrm{T,rel}}^2\\
    \pmb{T}_{\mathrm{T,rel}}\cdot({\pmb{1}_{\mathrm{T}}+\pmb{N}_{\mathrm{T}}\cdot\pmb{\Delta v}_{\mathrm{T,inc}}})\\
    \pmb{\tau}_{\mathrm{T,rel,i}}-\pmb{\tau}_{\mathrm{T,rel,r}}\cdot\left(\pmb{t}_{\mathrm{T}}\left(\pmb{m}_{\mathrm{T}}\right)\vert_{\nu}+\frac{\partial\pmb{t}\left(\pmb{m}_{\mathrm{T}}\right)}{\partial\pmb{m}_{\mathrm{T}}^{\mathrm{T}}}\vert_{\nu}\cdot\Delta\pmb{m}_{\mathrm{T}}\right)\\
    \pmb{m}_{\mathrm{T}}\\
    \pmb{n}_{\mathrm{T}}\\
    \pmb{T}_{\mathrm{T,rel,r}}\cdot\pmb{I}_{\mathrm{NT}}^{\mathrm{T}}\cdot\pmb{u}_{\mathrm{N,r}}-\pmb{T}_{\mathrm{T,rel,i}}\cdot\pmb{I}_{\mathrm{NT}}^{\mathrm{T}}\cdot\pmb{u}_{\mathrm{N,i}}-\pmb{u}_{\mathrm{T,a,r}}\\
    \pmb{T}_{\mathrm{T,rel,i}}\cdot\pmb{I}_{\mathrm{NT}}^{\mathrm{T}}\cdot\pmb{u}_{\mathrm{N,r}}+\pmb{T}_{\mathrm{T,rel,r}}\cdot\pmb{I}_{\mathrm{NT}}^{\mathrm{T}}\cdot\pmb{u}_{\mathrm{N,i}}-\pmb{u}_{\mathrm{T,a,i}}\\
    \pmb{G}_{\mathrm{TT}}\cdot\pmb{u}_{\mathrm{T,a,r}}-\pmb{B}_{\mathrm{TT}}\cdot\pmb{u}_{\mathrm{T,a,i}}-\pmb{i}_{\mathrm{T,a,r}}\\
    \pmb{G}_{\mathrm{TT}}\cdot\pmb{u}_{\mathrm{T,a,i}}+\pmb{B}_{\mathrm{TT}}\cdot\pmb{u}_{\mathrm{T,a,r}}-\pmb{i}_{\mathrm{T,a,i}}\\
    \pmb{T}_{\mathrm{T,rel,r}}\cdot\pmb{i}_{\mathrm{T,a,r}}+\pmb{T}_{\mathrm{T,rel,i}}\cdot\pmb{i}_{\mathrm{T,a,i}}-\pmb{i}_{\mathrm{T,r}}\\
    \pmb{T}_{\mathrm{T,rel,r}}\cdot\pmb{i}_{\mathrm{T,a,i}}-\pmb{T}_{\mathrm{T,rel,i}}\cdot\pmb{i}_{\mathrm{T,a,r}}-\pmb{i}_{\mathrm{T,i}}\\
    \pmb{i}_{\mathrm{T,r}}^2+\pmb{i}_{\mathrm{T,i}}^2-\pmb{i}_{\mathrm{T}}^2\\
    3\cdot\left(\pmb{U}_{\mathrm{N,r}}\cdot\pmb{I}_{\mathrm{NT}}\cdot\pmb{i}_{\mathrm{T,r}}+\pmb{U}_{\mathrm{N,i}}\cdot\pmb{I}_{\mathrm{NT}}\cdot\pmb{i}_{\mathrm{T,i}}\right)-\pmb{p}_{\mathrm{N0}}\\
    3\cdot\left(\pmb{U}_{\mathrm{N,i}}\cdot\pmb{I}_{\mathrm{NT}}\cdot\pmb{i}_{\mathrm{T,r}}-\pmb{U}_{\mathrm{N,r}}\cdot\pmb{I}_{\mathrm{NT}}\cdot\pmb{i}_{\mathrm{T,i}}\right)-\pmb{q}_{\mathrm{N0}}\\
    \pmb{u}_{\mathrm{N,r}}^2+\pmb{u}_{\mathrm{N,i}}^2-\pmb{u}_{\mathrm{N}}^2\\
    3\cdot\left(\pmb{u}_{\mathrm{N,r}}^{\mathrm{T}}\cdot\pmb{I}_{\mathrm{NT}}\cdot\pmb{i}_{\mathrm{T,r}}+\pmb{u}_{\mathrm{N,i}}^{\mathrm{T}}\cdot\pmb{I}_{\mathrm{NT}}\cdot\pmb{i}_{\mathrm{T,i}}\right)-P_{\mathrm{loss}}
\end{bmatrix}
\end{small}
\end{align}
With this system of equations, that only has real-valued polynomial functions of degree two and whose number of equations equals the number of variables, the SQCQP approach from \cite{Lev1} can be applied. As reasoned and undertaken in \cite{Lev1}, a distributed slack needs to be inserted additionally before inverting and convexifying. The resulting convexificated quadratically constrained quadratic  program is shown in Eq. \ref{eq:SQCQP_obj} to \ref{eq:SQCQP_Constraints2}.
\begin{align}\label{eq:SQCQP_obj}
\mathrm{min}\left(c_{\mathrm{loss}} \cdot P_{\mathrm{loss}}\left(\left[\pmb{m}_{\mathrm{T}}^{\mathrm{T}},\pmb{n}_{\mathrm{T}}^{\mathrm{T}},\mathrm{\Delta}\pmb{p}_{\mathrm{N}}^{\mathrm{T}},\mathrm{\Delta}\pmb{q}_{\mathrm{N}}^{\mathrm{T}}\right]\right)+\pmb{c}_{\mathrm{N}}^{\mathrm{T}} \cdot \mathrm{\Delta}\pmb{p}_{\mathrm{N}}\right)
\end{align}
s.t.
\begin{align}
\pmb{v}_{\mathrm{N,min}} \leq \pmb{v}_{\mathrm{N}}\left(\left[\pmb{m}_{\mathrm{T}}^{\mathrm{T}},\pmb{n}_{\mathrm{T}}^{\mathrm{T}},\mathrm{\Delta}\pmb{p}_{\mathrm{N}}^{\mathrm{T}},\mathrm{\Delta}\pmb{q}_{\mathrm{N}}^{\mathrm{T}}\right]\right) \leq \pmb{v}_{\mathrm{N,max}} \label{eq:SQCQP_Constraints1} \\
\pmb{i}_{\mathrm{T}}\left(\left[\pmb{m}_{\mathrm{T}}^{\mathrm{T}},\pmb{n}_{\mathrm{T}}^{\mathrm{T}},\mathrm{\Delta}\pmb{p}_{\mathrm{N}}^{\mathrm{T}},\mathrm{\Delta}\pmb{q}_{\mathrm{N}}^{\mathrm{T}}\right]\right) \leq \pmb{i}_{\mathrm{T,max}} \label{eq:SQCQP_Constraints2}
\end{align}
To enhance the speed of the sequential approach, the necessary relative MIPgap to be  is dynamically adjusted in each sequential step: In the first step, for which a heightened forecast error is to be expected due to the high state changings by the optimizer and the used approximations, it is set to a value of 10 \%. In the following steps it is set to the tenth of the MIPgap of the respective preceding step. Furthermore an dynamically adjusted time\hbox{-}limit is set for the optimizer, that starts with~128~s in the first step and doubles with each step. The total time for the duration one step can be longer, due to the convexification and further scripts.
\subsection{Particle Swarm Optimization Approach}\label{AdaptationPSO}
In the field of electric power system, optimization metaheuristics are used for solving a variety of optimization problems concerning system planning and operation \cite{ZhuOpt,STL,kaviani_optimal_2009,sharma2014comparative,sarstedt_application_2018,abido2002optimal,zhao_novel_nodate}. For solving SCOPF problems the PSO (cf. \cite{kennedy_particle_1995,eberhart_new_1995}) shows an appropriate convergence behavior and good performance characteristics by the adaptation of hyperparameters \cite{ZhuOpt,sharma2014comparative,sarstedt_application_2018}. The PSO algorithm used for the investigations within this paper is described in detail at \cite{Sar1}. For a better understanding of the specific adaptations for solving the SCOPF presented in section \ref{DefOPF} the main equations of the PSO and the general procedure are introduced. At the beginning of the iterative solution process the particle swarm consisting of $i=1,...,n$ individuals and $j=1,...,m$ control variables $x_{i,j}$ is initiated randomly within the corresponding flexibility limits of Eq. \ref{eq:Constraints1}, \ref{eq:Constraints4}, \ref{eq:Constraints5}, \ref{eq:ConstraintsTap1} and \ref{eq:ConstraintsTap2}. Analogously to the control variables, the velocity vector of the swarm particles $\pmb{v}_{i}$ is generated randomly for the initial iteration step $t=0$. For each swarm particle a power flow calculation based on the Newton Raphson algorithm is performed and the fitness value is evaluated by Eq. \ref{eq:SCOPF_obj} \cite{sharma2014comparative,ZhuOpt}. By this, the implementation of different control variables to the SCOPF problem is simple and is done by the integration of the variable to the power flow calculation algorithm. An additional punishment summand $g$ is added to Eq. \ref{eq:SCOPF_obj} for the non compliance of technical constraints. See \cite{Sar1} for details regarding the determination of the punishment summand $g$ \cite{ConstraintHandlingPSO,mohemmed_particle_nodate}. Based on the fitness values the global best swarm particle position $\pmb{p}_{\mathrm{gb}}$ of all particles as well as the individual best position of each particle $\pmb{p}_{\mathrm{b},i}$ during all iterations steps are updated. For the next iteration step the movement of the swarm and by this the change of the control variables are determined:
\begin{align} \label{Eq:Velo}
\pmb{v}_{i,t+1}=w\pmb{v}_{i,t}+c_{1}\pmb{r}_{1}\left(\pmb{p}_{\mathrm{b},i}-\pmb{x}_{i} \right)+c_{2}\pmb{r}_{2}\left(\pmb{p}_{\mathrm{gb},i}-\pmb{x}_{i}\right) \\
\mathrm{with} \ \pmb{r}_{1},\pmb{r}_{2}=(nx1), \pmb{v}_{i},\pmb{x}_{i}=(1xm) \\
\pmb{x}_{i,t+1}=\pmb{x}_{i,t}+k\pmb{v}_{i,t+1} \label{Eq:Position}
\end{align}
The acceleration coefficients $c_{1}$ and $c_{2}$ describe the social and the cognitive interactions within the swarm. The vectors $\pmb{r}_{1}$ and $\pmb{r}_{2}$ consist of random numbers in the interval of $[0,1]$ representing the stochastic nature of the PSO. 
The inertia $w$ indicates how the velocity of the swarm $\pmb{v}_{i,t+1}$ is affected by the current velocity $\pmb{v}_{i,t}$. The initial inertia $w$ decreases within the iteration process from $w_{\text{start}}$ to $w_{\text{end}}$ to guarantee at the beginning of the solution process a global and at the end ($t=t_{\mathrm{max}}$) a local search behavior of the swarm \cite{parsopoulos_particle_nodate, ZhuOpt}:
\begin{align}\label{eq:Inertia}
w=w_{\mathrm{start}}-t\left(\frac{w_{\mathrm{start}}-w_{\mathrm{end}}}{ t_{\mathrm{max}} }\right)
\end{align}
The constriction factor $k$ ensures a convergence of the swarm in a reliable solution of the optimization problem \cite{parsopoulos_particle_nodate, ZhuOpt}: 
\begin{align}\label{eq:ConstrictionFactor}
k=\frac{2}{\left\lvert 2-\psi-\sqrt{\psi^{2}-4\psi} \right\rvert}, \mathrm{with}\ \psi=c_{1}+c_{2}\geq4
\end{align}
The iterative solution process stops after the maximum iteration step $t_{\mathrm{max}}$ is reached. In the following the modifications of the classic PSO \cite{kennedy_particle_1995} approach regarding the solution of the SCOPF are introduced. 
Within the classic PSO only continuous variables are implemented to avoid a negative influence on the swarm behavior and to guarantee appropriate convergence. To consider integers for the in-phase and quadrature transformer tap sets the corresponding variables $z$ are considered as rounded values only during the evaluation of the swarm fitness \cite{zhao_novel_nodate}:   \begin{align} \label{Eq:Rounding}
x_{i,z}=\lfloor\frac{\lceil2x_{i,z}\rceil}{2} \rfloor
\end{align}
To avoid unfeasible solutions the movements of the particles is limited in front of the next iteration step within a set-to-limit operator \cite{STL}:
\begin{align} \label{Eq:STL}
x_{i,j}(x_{i,j}<x_{i,j,\mathrm{min}})&=x_{i,j,\mathrm{min}} \\
x_{i,j}(x_{i,j}>x_{i,j,\mathrm{max}})&=x_{i,j,\mathrm{max}}
\end{align}
The limitation of the swarm velocities is a common procedure to avoid alternating jumps between the control variable limits ($x_{i,j,\mathrm{min}}$, $x_{i,j,\mathrm{max}}$) and to guarantee a more detailed global solution search especially at the beginning of the PSO \cite{schutte_parallel_2004}. The speed coefficient $z$ is specified individually for each control variable (see Tab.~\ref{tab:Hyperparameters of the PSO}, active power redispatch $z_{\text{AR}}$, reactive power redispatch $z_{\text{RR}}$, quadrature voltage control transformer $z_{\text{TQ}}$, in-phase voltage control transformer $z_{\text{TIP}}$, voltage control power plants $z_{\text{VC}}$):
\begin{align} \label{Eq:VeloLimits}
v_{i,j}\left(|v_{i,j}|>v_{i,j,\mathrm{max}}\right)=\text{sgn}(v_{i,j})r_{3}v_{i,j,\mathrm{max}} \ \text{with}\\
v_{i,j,\mathrm{max}}=z(x_{i,j,\mathrm{max}}-x_{i,j,\mathrm{min}}), \ r_{3}=\text{rand}([0,...,1])
\end{align}
The movement of the swarm particles to unfeasible solutions and an accumulation of the swarm at the control variable limits are restricted by \cite{kennedy_particle_1995}:    
\begin{align} \label{Eq:VeloInvert}
v_{i,j}\left(x_{i,j}=x_{i,j,\mathrm{min}} \land v_{i,j}<0 \right)&=-v_{i,j} \\
v_{i,j}\left(x_{i,j}=x_{i,j,\mathrm{max}} \land v_{i,j}>0 \right)&=-v_{i,j}
\end{align}
At the beginning of the PSO the high inertia of the swarm enables a global solution search. As a result of the decreasing inertia over PSO iterations the search behavior becomes more local. For a better local solution search a mutation operator is introduced that manipulates a random control variable of each swarm particle \cite{ZhuOpt}:  
\begin{align} \label{Eq:Crazyness}
x_{i,r_{4}}=r_{5}v_{i,r_{4},\mathrm{max}} \ \forall \ i \in [1,...,n] \ \text{with}\\
r_{4}=\text{rand}([0,...,m]), r_{5}=\text{rand}([-1,...,1])
\end{align} 
The hyperparameters of the PSO (see Table \ref{tab:Hyperparameters of the PSO}) are selected manually based on references in the literature and experiences during the case study \cite{abido2002optimal,kennedy_particle_1995,sarstedt_application_2018,Sar1,polprasert_optimal_2016}. Multiple PSO runs $\lambda$ are performed in parallel due to the stochastic nature of the PSO and the possibility of local convergence \cite{schutte_parallel_2004}. 
\begin{table}[h]
\caption{Hyperparameters of the PSO}
\label{tab:Hyperparameters of the PSO}
\begin{tabular}{|c|c|c|c|c|c|c|c|}
\hline
$n$ & $t_{\mathrm{max}}$ & \begin{tabular}[c]{@{}c@{}}$c_{1}$\\ $c_{2}$\end{tabular} & $w_{\mathrm{start}}$ & $w_{\mathrm{end}}$ & \begin{tabular}[c]{@{}c@{}}$z_{\text{AR}}$\\ $z_{\text{RR}}$\end{tabular} & \begin{tabular}[c]{@{}c@{}}$z_{\text{TQ}}$ \\ $z_{\text{TIP}}$\\ $z_{\text{VC}}$ \end{tabular} & $\lambda$ \\ \hline
200 & 500 & 2 & 0.9 & 0.4 & 1/10 & 1/5 & 100 \\ \hline
\end{tabular}
\end{table}
\begin{table}[t]
\caption{Limits of the control variables and incremental voltage and in-phase change per transformer tap change}
\label{tab:ControlVariableLimits}
\begin{tabular}{|c|c|c|c|c|c|}
\hline
\begin{tabular}[c]{@{}c@{}}$\mathrm{\Delta}\pmb{p}_{\mathrm{N,min}}$ \\ $\mathrm{\Delta}\pmb{p}_{\mathrm{N,max}}$\\ $\mathrm{\Delta}\pmb{q}_{\mathrm{N,min}}$ \\ $\mathrm{\Delta}\pmb{q}_{\mathrm{N,max}}$ \end{tabular} & \begin{tabular}[c]{@{}c@{}}$ \pmb{v}_{\mathrm{N,min}}$\\ $\pmb{v}_{\mathrm{N,max}}$\end{tabular}  & \begin{tabular}[c]{@{}c@{}}$ \pmb{m}_{\mathrm{min}}$\\ $ \pmb{m}_{\mathrm{max}}$\end{tabular} & \begin{tabular}[c]{@{}c@{}}$ \pmb{n}_{\mathrm{min}}$\\ $ \pmb{n}_{\mathrm{max}}$\end{tabular} &  $\Delta \pmb{V}_{\mathrm{inc}}$&  $\Delta \pmb{\phi}_{\mathrm{inc}}$\\ \hline
\begin{tabular}[c]{@{}c@{}}see \cite{DATASET}\\ see Fig. \ref{Fig.Flex}\end{tabular}& \begin{tabular}[c]{@{}c@{}}0.9 p.u.\\ 1.1 p.u.\end{tabular} & \begin{tabular}[c]{@{}c@{}}-10\\ 10 \end{tabular} & \begin{tabular}[c]{@{}c@{}}-10\\ 10 \end{tabular} & 0.25 \% &  1° \\ \hline
\end{tabular}
\end{table}
\section{Case-Study}\label{CaseStudiesGeneral}
\begin{figure}[b!]
    \centering
    \includegraphics[width=220pt]{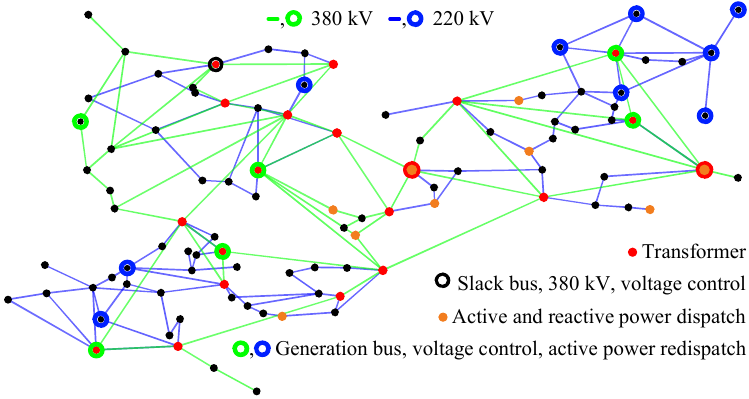}
    \caption{IEEE 118-bus system with control variables}
    \label{fig:NetzPlot}
\end{figure}

The SQCQP and the PSO approach presented in sections~\ref{AdaptationSQCQP} and \ref{AdaptationPSO} are both implemented in MathWorks Matlab. The simulations are performed on computers with a 2.7~GHz QuadCore and 16 GB RAM. For the comparison of the SQCQP and the PSO approach an adaptation of the IEEE 118-bus transmission grid regarding German transmission grid characteristics (see Fig. \ref{fig:NetzPlot}, cf. \cite{Aachen}) is used. A MathWorks Matlab dataset of the grid model and the results of the case study for the SQCQP and the best PSO run are available at \cite{DATASET}.
The limits of the control variables as well as the incremental voltage and phase change per transformer tap change $n$ and $m$ are given in Tab. \ref{tab:ControlVariableLimits}. The costs for grid losses $c_{\mathrm{loss}}$ are set to~1 while the costs for active and reactive redispatch $c_{\mathrm{N}}$ are set to 0. Bus number 63 is selected as slack and the voltage phase is set to 0°. The initial grid losses with the voltage control from~\cite{Aachen} are $P_{\mathrm{loss},0}=189.90 \text{MW}$.  
The investigations within the case-study are divided into three scenarios. In scenario 1 operational degrees of freedom are only provided by power plants that are able to contribute to active and reactive power redispatch and to perform the voltage control at the generation buses, respectively. In scenario 2 additional flexibilities are provided by the incremental in-phase voltage control of the transformers. In scenario 3 additionally the incremental quadrature voltage control of the transformers is considered. 
\subsection{Scenario 1}\label{Case1}
\begin{figure}[t!]
    \centering
    \includegraphics[width=230pt]{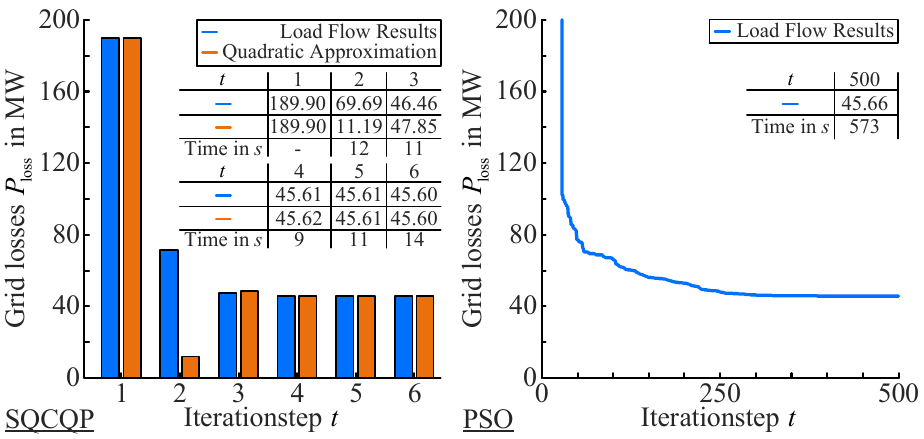}
    \caption{Results for scenario 1}
    \label{fig:ErgebnisVerlauf}
\end{figure}

In Fig. \ref{fig:ErgebnisVerlauf} the convergences of the SQCQP and the PSO for scenario 1 are presented. The convexificated quadratic approximations of the SQCQP approximate the non-linear system behavior well except for the second sequential step (see Fig. \ref{fig:ErgebnisVerlauf} left). 
\begin{figure}[b!]
    \centering
    \includegraphics[width=220pt]{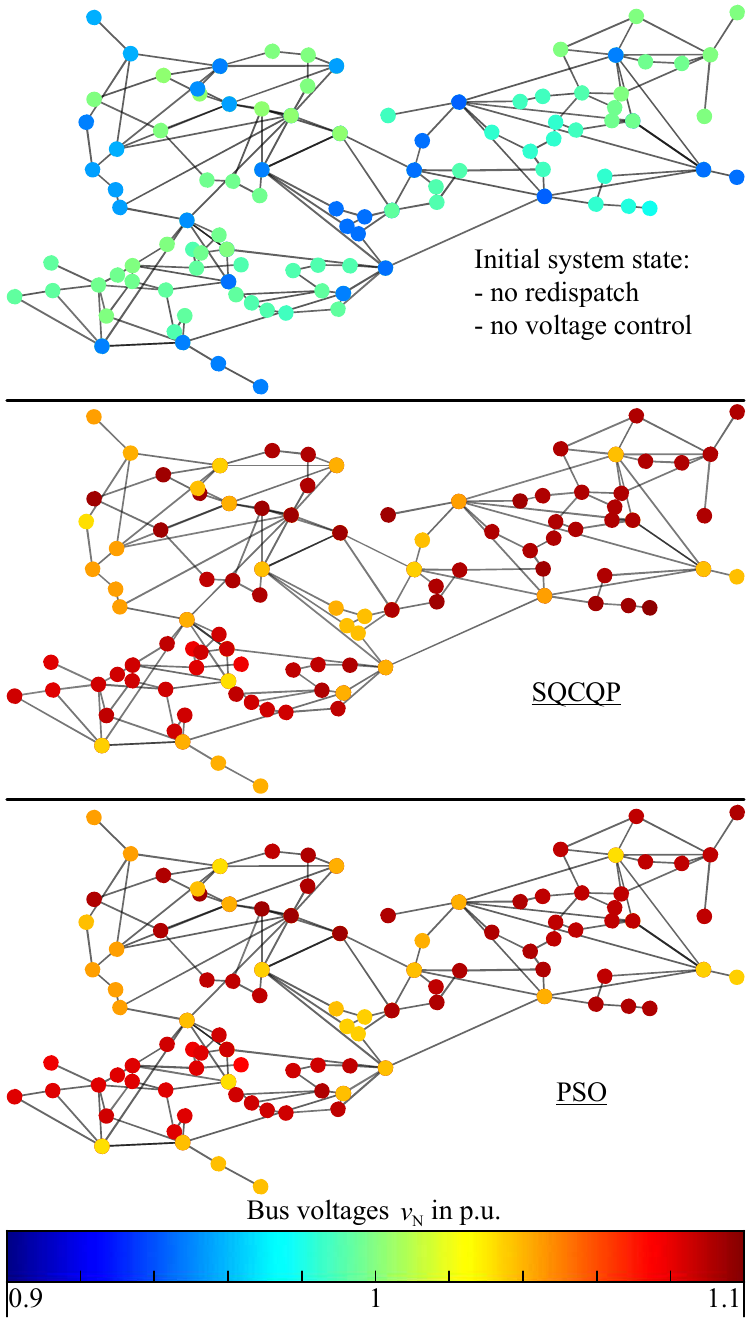}
    \caption{Initial and resulting nodal voltages in scenario 1}
    \label{fig:ErgebnisSpannung}
\end{figure}
The SQCQP finds a slightly better solution ($\Delta P_{\mathrm{loss}}=0.06 \ \mathrm{MW}$) and the computation is faster ($\Delta t=516 \ \mathrm{s}$). Within the $\lambda$ PSO runs the average result is $P_{\mathrm{loss,av}}=46.34 \ \mathrm{MW}$ and the worst result is ${P_{\mathrm{loss,w}}=46.83 \ \mathrm{MW}}$. In contrast to that, another advantage of the SQCQP in scenario 1 is the reproducibility of the results.\\
The exclusive monetarization of grid losses in the objective function leads to significant active and reactive power redispatches and an increase of the control voltages at the generation buses (see \cite{DATASET}). For the resulting absolute values of the nodal voltages $\pmb{v}_{\mathrm{N}}$ in Fig. \ref{fig:ErgebnisSpannung} just small deviations between the SQCQP and the PSO are identified. This observation also applies to the following case studies, so the evaluation of the voltages is not presented again.
\subsection{Scenario 2}\label{Case2}
In scenario 2 again the SQCQP gives a slightly better result then the PSO ($\Delta P_{\mathrm{loss}}=0.02 \ \mathrm{MW}$) in a lower computation time ($\Delta t=418 \ \mathrm{s}$). The scattering of the PSO results is reduced compared to scenario 1 with an average result of $P_{\mathrm{loss,av}}=45.54 \ \mathrm{MW}$ and the worst~result~${P_{\mathrm{loss,w}}=45.61 \ \mathrm{MW}}$. The differences between the results of scenario 1 and 2 are small due to an already high utilization of the voltage limits based on reactive power optimization and voltage control in scenario 1 (see Fig. \ref{fig:ErgebnisSpannung}) as well as the dominating influence of the in-phase voltage control on the same voltage magnitudes.
\begin{figure}[t!]
    \centering
    \includegraphics[width=230pt]{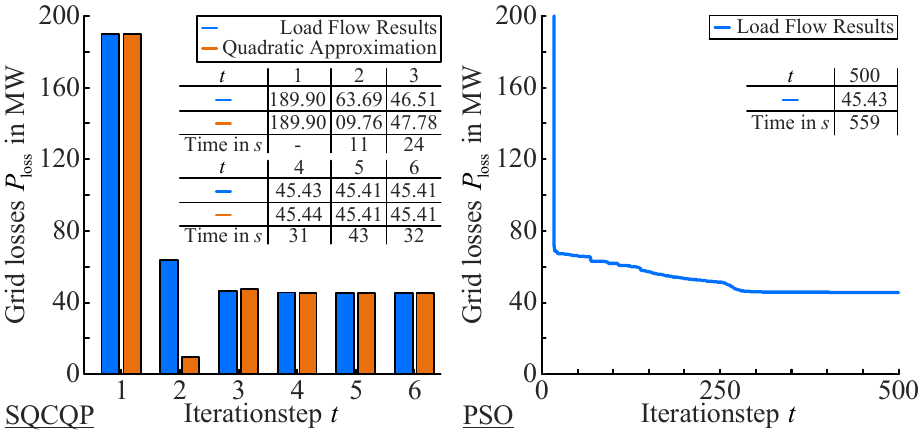}
    \caption{Results for scenario 2}
    \label{fig:Case2}
\end{figure}
\begin{figure}[t!]
    \centering
    \includegraphics[width=230pt]{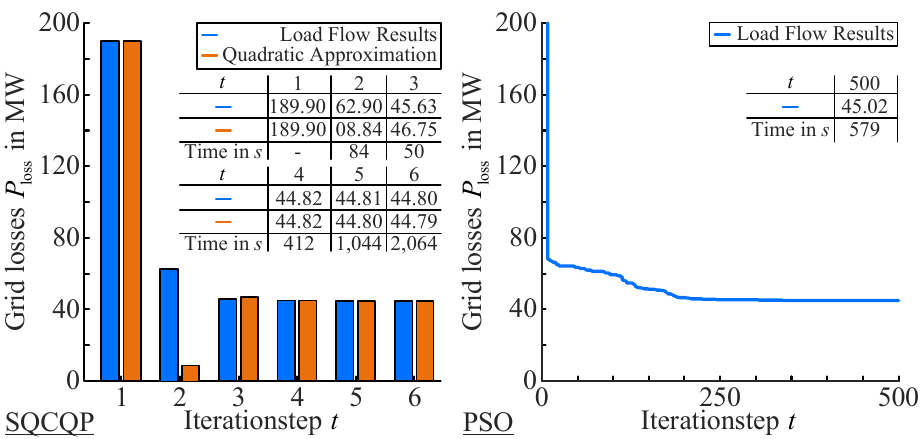}
    \caption{Results for scenario 3}
    \label{fig:Case3}
\end{figure}
\subsection{Scenario 3}\label{Case3}
\begin{figure}[t!]
    \centering
 \begin{flushleft}
\includegraphics[width=230pt]{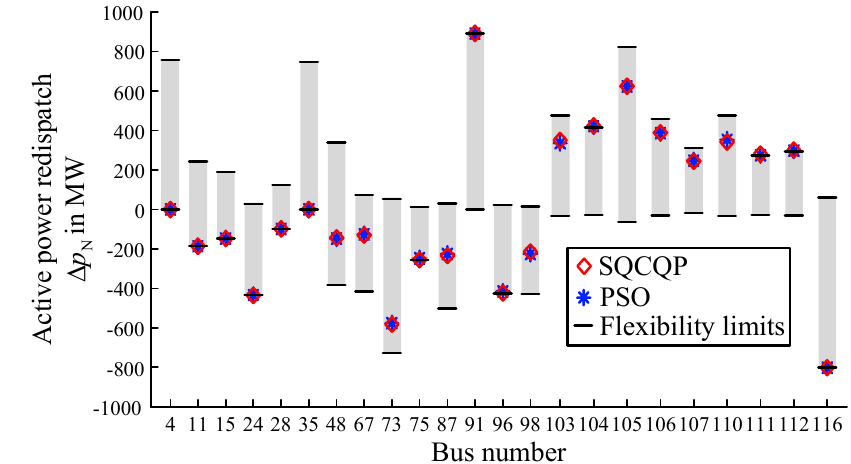}\\[-1pt]
\includegraphics[width=230pt]{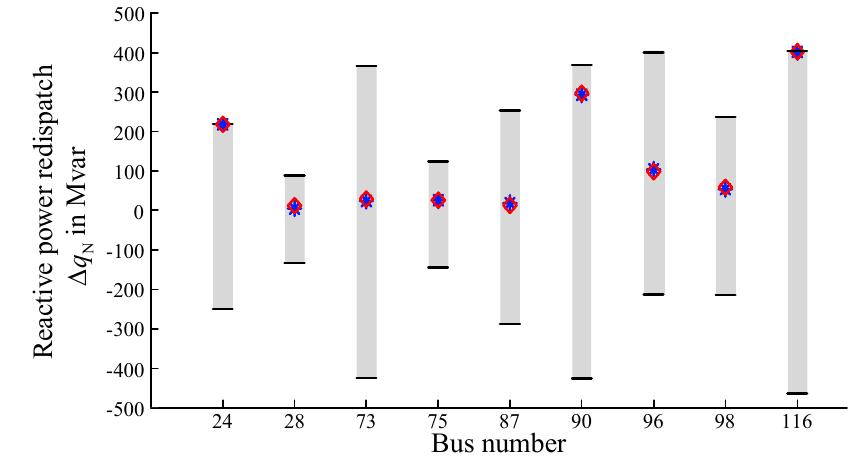}\\[-1pt]
\includegraphics[width=230pt]{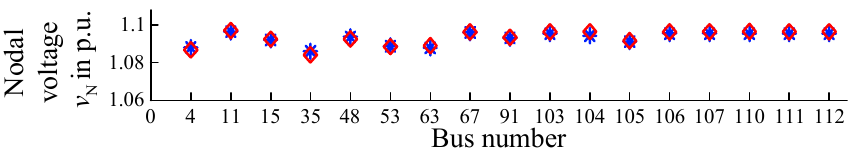}\\[-1pt]
\includegraphics[width=230pt]{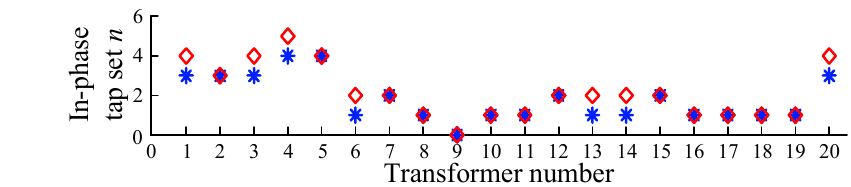}\\[-1pt]
\includegraphics[width=230pt]{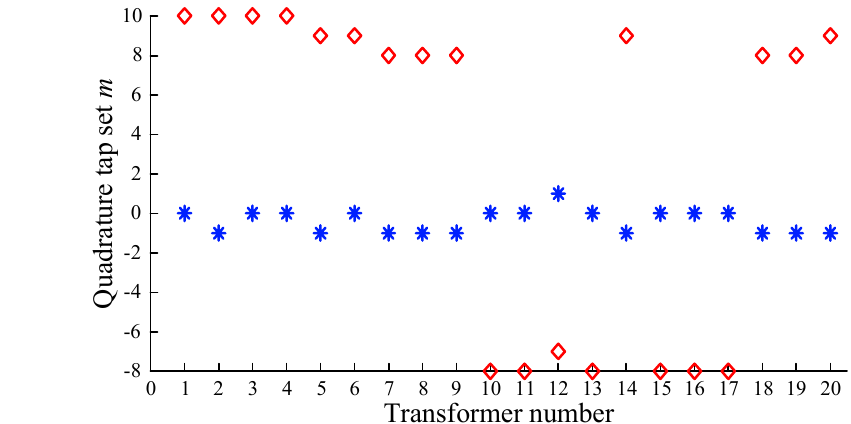}
\caption{Utilization of control variables in scenario 3}
\label{Fig.Flex}
\end{flushleft}
\end{figure}
In scenario 3 the additional consideration of the quadrature voltage control significantly lowers the objective value. The impact is higher than that of the in-phase voltage control in scenario 2. This can be reasoned with the dominating influence of the quadrature voltage control on the distribution of the active power flows in the grid and consequently on the losses. The SQCQP reduces the grid losses more then the PSO ($\Delta P_{\mathrm{loss}}=0.23 \ \mathrm{MW}$), but needs significantly more time, despite the time limits defined in section \ref{AdaptationSQCQP}. Compared to scenario 2 the scattering of the PSO results increases slightly with an average result of ${P_{\mathrm{loss,av}}=45.14 \ \mathrm{MW}}$ and the worst result $P_{\mathrm{loss,w}}=45.30 \ \mathrm{MW}$. In general, both approaches utilize the flexibility potentials of the control variables in the same way (see Fig. \ref{Fig.Flex}). Significant differences only arise for the in-phase voltage control and especially for the quadrature voltage control of the transformers.              

\section{Conclusion and Outlook}
The flexibilities in  classical Security Constrained Optimal Power Flows (SCOPF) are often only represented by active and reactive power redispatch and voltage control measures of thermal power plants or aggregated renewable energy resources. Within this paper, incremental in-phase and quadrature voltage controlled transformers are considered as additional, mixed-integer control variables. For the solution of this more complex SCOPF problem two different powerful optimal power flow solvers, namely a Sequential Quadratic Constrained Quadraic Programming (SQCQP) approach and a modified particle swarm optimization (PSO), are introduced and compared with each other regarding the quality of the results and the computation time within a case-study of an adapted IEEE~118\hbox{-}bus~system. Within the three investigation scenarios an increasing number of control variables are considered. At scenario 1 only active and reactive power redispatch and voltage control measures are implemented. In scenario 2 and 3 additional the incremental in-phase and quadrature voltage control of the transformers are taken into account.\\
The results of the case study can be concluded as follows: 
In comparison of the influence on the objective value, it can be seen, that the in-phase voltage control as well as the quadrature voltage control only have a smaller impact on the objective value compared to the reactive and specially the active power redispatch. The advantages of the voltage controllers are likely to become visible, when current and voltage congestions play an important role in the dataset. The results of the scenario~1 already show differences and specific advantages of the SQCQP and the PSO in finding an optimal solution of the SCOPF: Both algorithms nearly reach the same objective value, but the PSO takes a longer computational time. So the advantages of analytical approaches for continuous functions can be seen. The results of scenario~2, that additionally considers incremental in-phase voltage control of transformers, again reveal nearly equality of the gained objective value and an advantage of the SQCQP with respect to the summed computation time of all sequential steps. The results of the scenario 3, that additionally considers in-phase and quadrature voltage control of transformers, again reveal nearly equality of the gained objective value that can be also identified by the comparison of the control variables in Fig. \ref{Fig.Flex}. The convergence of both approaches to nearly the same control variable utilization lead to the assumption that the solution of the optimization problem has no pareto optimum. In contrast to scenario 1 and 2 a significant advantage of the PSO with respect to computation time arise in scenario~3. Comparing all scenarios the computation times of the PSO are constant because it scales only with the number of power flow calculations and the performance of the computer in case of background processes. In contrast to that, the integration of the incremental flexibilities to the SQCQP leads to a more complex problem and by this to an increased computation time ($\Delta t=3075 \ \mathrm{s}$). It was observed that the SQCQP has stuck in lowering the MIPgap without finding new solutions, but trying to increase the lower bound, when considering quadrature voltage control. Equivalent or slightly different solutions of the SCOPF are suspected in this context and this aspect as well as performance and accuracy improvements of the SQCQP will be part of future research. Similar future goals hold for the hyperparameter tuning of the PSO (e.g. swarm size, see Tab. \ref{tab:Hyperparameters of the PSO}) or the introduction of a sequential solution process. The optimization approaches and investigations are aimed to be extended by network configurations, flexibility potentials of underlying voltage levels or costs for redispatch by linking the simulation to a market simulation.

\bibliography{Bibliothek}

\end{document}